\documentclass[aps,pra,twocolumn,showpacs,superscriptaddress]{revtex4} 
\usepackage{graphicx}
\usepackage{amsmath}
\usepackage{amssymb}
\usepackage{epstopdf}
\usepackage{amsfonts}
\usepackage{dcolumn}
\usepackage{bigints}
\usepackage{xcolor}
\usepackage[normalem]{ulem}
\usepackage{natbib}
\begin{document}
\title{Structural properties of $^4$He$_N$ ($N=2-10$) clusters for different potential models
at the physical point and at unitarity}
\author{A. J. Yates}
\address{Homer L. Dodge Department of Physics and Astronomy,
  The University of Oklahoma,
  440 W. Brooks Street,
  Norman,
Oklahoma 73019, USA}
\address{Center for Quantum Research and Technology,
  The University of Oklahoma,
  440 W. Brooks Street,
  Norman,
Oklahoma 73019, USA}
\author{D. Blume}
\address{Homer L. Dodge Department of Physics and Astronomy,
  The University of Oklahoma,
  440 W. Brooks Street,
  Norman,
Oklahoma 73019, USA}
\address{Center for Quantum Research and Technology,
  The University of Oklahoma,
  440 W. Brooks Street,
  Norman,
Oklahoma 73019, USA}
\date{\today}

\begin{abstract}
Since the $^4$He dimer supports only one weakly bound state with an average interatomic distance
much larger than the van der Waals length and no deeply bound states, $^4$He$_N$ clusters
with $N>2$
are a paradigmatic model system with which to explore foundational concepts such
as large 
$s$-wave scattering length 
universality, van der Waals universality, Efimov physics,
and effective field theories. This work presents structural properties such as the
pair and triple distribution functions, the hyperradial density,  the probability to find the $N$th particle at a 
given distance from the center of mass of the other $N-1$ atoms, and selected contacts.      
The kinetic energy release,  which can be measured via Coulomb explosion in dedicated
size-selected  molecular beam experiments---at least for small $N$---, is also presented.
The structural properties are determined for three different realistic $^4$He-$^4$He interaction potentials and 
contrasted with those for an effective low-energy potential model from the literature
that reproduces the 
 energies of $^4$He$_N$ clusters in the ground state for $N=2$ to $N=\infty$ at the 
 $\gtrsim 95$~\% level
with just four input parameters.  
 The study is extended to unitarity (infinite $s$-wave scattering length)
by artificially weakening the interaction potentials.
In addition to contributing to the characterization of small bosonic helium quantum droplets,
our study provides insights into the effective low-energy theory's predictability of
various structural properties.
\end{abstract}
\maketitle

\section{Introduction}
\label{sec_introduction}
Bosonic helium droplets, i.e., clusters consisting of a finite number of $^4$He (helium-4) atoms, have captivated physicists' interests over many decades~\cite{PhysRevB.34.4571,
PhysRevLett.50.1676,
PhysRevA.47.4082,
PhysRevLett.38.341,
doi:10.1063/1.450912,
doi:10.1080/01442359409353290,
doi:10.1063/1.473501,
PhysRevA.54.394,
doi:10.1063/1.2140723,
doi:10.1080/00268976.2013.802039,
PhysRevB.52.10405,
PhysRevB.52.1193,
doi:10.1063/1.1310608,
doi:10.1063/1.470771,
doi:10.1126/science.266.5189.1345,
doi:10.1063/1.470772,
PhysRevLett.85.2284,
PhysRevA.67.042706,
PhysRevA.85.062505,
PhysRevA.85.022502,
doi:10.1063/1.458943,
PhysRevA.64.042514,Nielsen_1998,doi:10.1063/1.482027,Blume2015,
doi:10.1126/science.aaa5601,
Kolganova2011,
2016,
PhysRevA.86.012502,
PhysRevA.90.032504,
PhysRevA.96.040501,
doi:10.1063/1.481404,
PhysRevA.102.063320,
PhysRevA.101.010501}.
They provide a bridge
between the microscopic and macroscopic worlds, with the helium dimer
being bound by just 1.6~mK~\cite{PhysRevLett.104.183003,Zeller14651} 
and the binding energy per particle reaching
about 7~K in bulk liquid helium-4~\cite{RevModPhys.67.279}. 
Mesoscopic helium-4 droplets are essentially incompressible
and
a subset of their properties are captured accurately by a ``bare bone'' liquid drop model, which contains a volume term,
a surface term, and two additional terms that are treated as fitting
parameters~\cite{PhysRevB.52.1193}. 
We note that liquid drop models that contain volume, surface, 
Coulomb, pairing, and asymmetry terms provide a starting point for understanding key properties of
nuclei, including the stability of highly-deformed nuclei and nuclear 
fission~\cite{PhysRevC.67.044316}.
The roton minimum, the smoking gun of superfluid bulk helium-4~\cite{PhysRev.102.1189,PhysRev.121.1266}, 
has been found theoretically
to first emerge for $N \approx 60$ atoms~\cite{doi:10.1063/1.459525}, 
motivating the term microscopic superfluidity. Experimentally, the prediction was verified by
embedding a small helium-4 cluster of varying size into a much larger helium-3 
droplet~\cite{doi:10.1126/science.279.5359.2083}.
Large helium-4 droplets with more than about $N=1,000$ atoms, in turn, have been employed as micro-laboratories with which
to capture, cool, and equilibrate impurities of varying size, from
single atoms to proteins~\cite{https://doi.org/10.1002/anie.200300611,MAURACHER20181,PhysRevLett.105.133402,doi:10.1063/1.1359707,Stienkemeier_2006,doi:10.1063/1.1418746,
doi:10.1146/annurev.physchem.49.1.1}.
  
  This paper provides a detailed analysis of various observables of 
  small pristine $^4$He$_N$ clusters, $N=2-10$, that interact either through a sum
  of realistic 
   two-body potentials~\cite{doi:10.1063/1.438007,doi:10.1063/1.4712218,PhysRevLett.74.1546}
   or an effective low-energy 
  model potential that includes 
  two- and three-body terms~\cite{PhysRevA.102.063320}. 
  Emphasis is placed on structural properties, including  the long-distance
  tail  and the short-distance correlations (on the scale of the van der Waals
  length $r_{\text{vdW}}$) of the pair, Jacobi, and hyperradial distribution functions.
  The long-distance tails are expected to be governed by the effective low-energy
  model, i.e., the large-distance fall-off should
  be fully governed by the binding
  energy, which was matched for $N=2-4$ when constructing the effective low-energy model
 and reproduces the exact binding energies at the $\gtrsim 95$~\%
 level  for $N>4$~\cite{PhysRevA.102.063320}.  
 
 The short-distance correlations are expected to be governed by the two-body wave function
 for an attractive van der Waals potential, i.e., a potential with $-C_6/r^6$ tail~\cite{PhysRevLett.112.105301,PhysRevA.90.022106}.
 For helium clusters that have been ``artificially'' scaled to the unitary point, this has previously been confirmed through dedicated
 calculations~\cite{PhysRevLett.112.105301,PhysRevA.90.022106}. 
 For the physical point, this is demonstrated, to the best of our knowledge, 
  for the first time in this work. 
 The collapse of the pair, triple, and higher-order distribution functions for interatomic distances
  around
 $r \approx r_{\text{vdW}}$ was noted in the literature, motivating the adaption of the $n$-body Tan contact ($n=2,3,\cdots$)~\cite{TAN20082987,TAN20082971,TAN20082952,PhysRevA.86.013626,PhysRevA.86.053633,PhysRevLett.106.153005} 
 to quantum clusters that exhibit weak universality (two-body potentials with van der Waals tail 
 and strongly repulsive hard-wall like short-distance repulsion)~\cite{PhysRevA.101.010501}.
 This work compares the two-body contact with results from the literature and 
 additionally introduces a $(2+1)$ contact.
 We note that short-distance correlations, or generalizations of the Tan contact from zero- to finite-range
 interactions, are also being actively investigated in 
 nuclei~\cite{PhysRevLett.114.012501,PhysRevC.92.054311,MILLER2018442,doi:10.1126/science.1256785,Cruz-Torres2021}.
 
 Since the effective low-energy potential 
 model does not ``know'' about $r_{\text{vdW}}$, the short-distance or high-energy
 correlations are not captured by that model despite the fact that
 the short-distance physics is universal, i.e., governed by 
 the $s$-wave scattering length $a_s$
 and the van der Waals length $r_{\text{vdW}}$. Thus, 
 the development of a low-energy van der Waals theory is highly desirable.
While this is beyond the scope of the present paper,
we note that first steps in this direction were recently taken~\cite{PhysRevA.104.023306}. 
Our aim in the present work is to provide a careful analysis of the different ``universality regimes,''
providing a comprehensive study of the structural properties of pristine helium-4 clusters at the physical point and at unitarity.

Many naturally occurring systems are characterized by competing length or energy scales. 
The helium-helium interaction is unique in that its naturally occurring $s$-wave scattering length 
$a_s$ is more
than an order of magnitude larger than its effective range. Even though $a_s$ is large, it is not infinitely large.
The regime where $a_s$ goes to infinity and the effective range goes to zero has been
studied quite extensively, not only in the context of bosons but also in the context of fermions. Correspondingly,
a comparison of the behaviors of small helium clusters at the physical point and those of helium clusters at unitarity
provides insights into developing universal van der Waals theories.
The lessons learned have importance beyond atomic droplets---some findings carry over to
the nuclear chart, with the weakly bound triton, alpha-particle, and halo-nuclei playing loose analogs of 
weakly-bound atomic clusters.

The remainder of this paper is organized as follows.
Section~\ref{sec_theory} introduces the system Hamiltonian, 
describes the numerical techniques employed to solve the time-independent Schr\"odinger
equation for clusters consisting of up to $N=10$ atoms, and defines  several
structural observables of interest.
Section~\ref{sec_results} presents and interprets our results. 
Connections with the literature are 
established throughout.
Finally, Sec.~\ref{sec_conclusion} summarizes and offers an outlook.

\section{Theoretical background}
\label{sec_theory}

\subsection{Hamiltonian}
Each  $^4$He atom is treated as a point particle with position vector 
$\vec{r}_j$ ($j=1,\cdots,N$) and mass $m$~\cite{footnote_mass}.
The non-relativistic $N$-atom Hamiltonian $\hat{H}$ reads
\begin{eqnarray}
\hat{H} = \sum_{j=1}^N \frac{-\hbar^2}{2m} \vec{\nabla}^2_{\vec{r}_j} + V_{\text{int}}({r}_{1,2},\cdots,{r}_{N-1,N}).
\end{eqnarray}
The interaction potential $V_{\text{int}}$ depends on the interatomic distances $r_{j,k}$, where
$r_{j,k}$ is equal to $|\vec{r}_j-\vec{r}_k|$.
We consider two different classes of interaction potentials $V_{\text{int}}$, referred to as Model~I and Model~II. For Model~I,
$V_{\text{int}}$  consists of a sum over 
two-body Born-Oppenheimer potentials $V_{\text{BO}}(r_{j,k})$, which have a repulsive core 
at small interatomic distances due to the electron repulsion and Pauli exclusion principle and
an attractive van der Waals tail with leading order term $-C_6/(r_{j,k})^6$,
\begin{eqnarray}
V_{\text{int}}(r_{1,2},\cdots,r_{N-1,N})=\sum_{j=1}^{N-1} \sum_{k>j}^N V_{\text{BO}}(r_{j,k}).
\end{eqnarray}
Calculations for Model~I are performed for three variants; specifically, 
we consider the Born-Oppenheimer potentials by Aziz {\em{et al.}}~\cite{doi:10.1063/1.438007} (HFD-HE2 potential, Model~IA), 
Cencek {\em{et al.}}~\cite{doi:10.1063/1.4712218} (CPKMJS potential, Model~IB), and
Tang {\em{et al.}}~\cite{PhysRevLett.74.1546}  (TTY potential, Model~IC).

For Model~II, $V_{\text{int}}$ is taken to be the effective low-energy potential
developed by Kievsky {\em{et al.}}~\cite{PhysRevA.102.063320,PhysRevA.96.040501},
\begin{eqnarray}
V_{\text{int}}(r_{1,2},\cdots,r_{N-1,N})=\sum_{j=1}^{N-1} \sum_{k>j}^N V_{2,G}(r_{j,k})+ \nonumber \\
\sum_{j=1}^{N-2} \sum_{k>j}^{N-1}\sum_{l>k}^N V_{3,G}(R_{j,k,l}),
\end{eqnarray}  
where $V_{2,G}(r_{j,k})$
and $V_{3,G}(R_{j,k,l})$ denote two- and three-body Gaussian potentials,
\begin{eqnarray}
V_{2,G}(r_{j,k}) = w_0 \exp \left[ -(r_{j,k}/r_0)^2 \right]
\end{eqnarray}
and
\begin{eqnarray}
V_{3,G}(R_{j,k,l}) = W_0 \exp \left[ -(R_{j,k,l}/R_0)^2 \right]
\end{eqnarray}
with
\begin{eqnarray} 
R_{j,k,l}^2 = \frac{1}{9} (r_{j,k}^2 + r_{j,l}^2 + r_{k,l}^2).
\end{eqnarray}
Reference~\cite{PhysRevA.102.063320} adjusted the parameters $w_0$, $r_0$, $W_0$, and $R_0$ 
such that the low-energy Hamiltonian reproduces
the ``exact'' two-body $s$-wave scattering length
and the $N=2,3$, and $4$ ground state energies of the realistic HFD-HE2 Born-Oppenheimer
potential (Model~IA)~\cite{typo_gaussian}.
While the two-body potential $V_{2,G}$ is
%purely 
attractive 
%at short 
for all
distances $r_{j,k}$ (i.e., $w_0$ is negative), the three-body potential is
purely repulsive 
%at small 
for all
hyperradii $R_{j,k,l}$ (i.e., $W_0$ is positive). The effective three-body repulsion
``counteracts'' the strong short-distance attraction of $V_{2,G}$. For $W_0=0$, the Gaussian interaction model yields a ground state energy that scales
as $N^2$~\cite{PhysRevA.90.032504,PhysRevA.92.033626}. 
The finite repulsive three-body term changes the scaling 
for $N$ up to about $10$ to approximately $N$~\cite{von_Stecher_2010,PhysRevA.92.033626,PhysRevA.96.040501,PhysRevA.102.063320}, in agreement with what is being 
observed for the realistic interaction potentials (Model~IA, Model~IB, and Model~IC).

The $^4$He-$^4$He potential is characterized by a large $s$-wave scattering, i.e., a scattering length
that is
about $35$ to
$45$ times larger than the van der Waals length $r_{\text{vdW}}$, $r_{\text{vdW}} \approx 5$~$a_0$
(the exact ratio depends on the interaction potential); see, e.g., Refs.~\cite{PhysRevA.67.042706,doi:10.1063/1.469978}.
We employ the definition $r_{\text{vdW}} = (m C_6/\hbar^2)^{1/4}/2$.
The scale separation is a key requirement for the emergence of Efimov physics in the
three-body sector~\cite{BRAATEN2006259,BRAATEN2007120,Naidon_2017}. 
Indeed, the first excited state of the $^4$He trimer, which has been probed 
experimentally~\cite{doi:10.1126/science.aaa5601}, has been 
identified as an essentially pure Efimov state, i.e., a state that can be described with high accuracy by just two input parameters (the $s$-wave scattering length and a three-body 
parameter)~\cite{PhysRevLett.38.341,doi:10.1063/1.450912,Kolganova2011,PhysRevA.86.012502,PhysRevA.54.394,Nielsen_1998,BRAATEN2006259,Blume2015}. 
In contrast, finite-range effects enter into the description of the $^4$He trimer ground 
state~\cite{PhysRevA.79.022702,doi:10.1126/science.aaa5601,Blume2015}.
Despite of this, the low-energy model (Model~II) reproduces the ground state energies of 
$^4$He$_N$ clusters with $N=2$ to $N=\infty$ remarkably well (i.e., at the  $\gtrsim 95$~\% 
level)~\cite{PhysRevA.102.063320}.

To investigate the regime where the 
two-body $s$-wave scattering length $a_s$
diverges, we follow the literature and scale 
$V_{\text{int}}$ 
by $\lambda$ ($\lambda <1$); we do this for Model~IA and Model~IB, 
choosing $\lambda$ such that the $s$-wave scattering length of $V_{\text{BO}}$ is infinitely large.
For Model~IA, we use $\lambda= 0.9792445$~\cite{PhysRevA.102.063320}; 
because our mass is slightly different than
that used in Ref.~\cite{PhysRevA.102.063320}, the resulting scattering length is large 
but not infinitely large ($1/a_s \approx 10^{-5}$~$a_0^{-1}$). 
For Model~IB, we use $\lambda=0.9713665$~\cite{Blume2015}, 
resulting in $1/a_s \approx 10^{-7}$~$a_0^{-1}$.
We note that the scaling 
changes the van der Waals length and effective range of $V_{\text{BO}}$ only slightly~\cite{effective_range}.
For Model~II, we use again the parameters from Kievsky {\em{et al.}}~\cite{PhysRevA.102.063320}; 
while $r_0$ and $R_0$ remain the same as at the physical
point, $|w_0|$ and $W_0$ are, respectively, 
slightly smaller and slightly larger at unitarity than at the physical point.
As stated, the scaling factors $\lambda$ and the parameters of the effective low-energy model 
are taken from the literature.
Since the $m$ values employed in the literature differ, the resulting scattering lengths
are very large but not infinitely large; 
we emphasize that this does not impact the conclusions of the paper.

\subsection{Monte Carlo techniques}

Our $N \ge 3$ results are obtained by the diffusion Monte Carlo (DMC) method~\cite{doi:10.1119/1.18168,Austin2012,doi:10.1063/1.465195},
which
yields
the ground state energies and structural properties of the ground state. 
The DMC method with importance sampling
utilizes a guiding or trial wave function $\psi_T$ that is optimized using the
variational Monte Carlo (VMC) technique~\cite{doi:10.1063/1.460853}.
The nodeless guiding or trial wave function
$\psi_T$, which depends on a set of non-linear variational parameters $\vec{p}$, is optimized by minimizing the
energy expectation value, which is evaluated stochastically using Metropolis sampling~\cite{doi:10.1063/1.1699114}.
If the walker number is sufficiently large and the imaginary time step $\tau$ sufficiently small,
the DMC energies are, within statistical uncertainties, exact. 
We use  between $2,000$ and $5,000$ walkers for all $N$ considered. 
The energy 
is calculated using the growth 
estimator  and the mixed estimator,
yielding the growth energy $E_g$ and the mixed energy $E_m$, respectively. 
For the calculations reported in Tables~\ref{table1} and \ref{table1_unit}, 
the two estimators yield consistent energies, i.e., the distribution of the energies
and 
errors
are consistent with the fact that the 
errors indicate a 68~\% confidence interval. 
For $N=2$, a statistically significant time step dependence
is observed (see 
%Fig.~\ref{SI_fig1} 
Fig.~S1
in the Supplemental Information~\cite{supplemental_material} for details); the caption
of 
Table~\ref{table1} 
reports the
extrapolated zero imaginary time step energies $E_g$.
 The time step dependence for $N=3$ is smaller than for 
 $N=2$ but still, at least for Models~IA-IC, statistically significant
 (see 
 %Fig.~\ref{SI_fig2} 
 Fig.~S2
 of the Supplemental Material).
 Correspondingly, Tables~\ref{table1} and \ref{table1_unit} report 
 extrapolated zero imaginary time step growth energies $E_g$.
 For $N>3$, the time step dependence is 
 estimated to be smaller than $0.5$~\% and Tables~\ref{table1} and \ref{table1_unit} report 
growth energies that are obtained for
a fixed imaginary time step ($\tau$ between $200$ and $400$~a.u., where 
``a.u.'' stands for ``atomic units'').

\begin{widetext}

\begin{table}
\begin{tabular}{c|cccc|cc}
$N$ & $E_{\text{HFD-HE2}}$$^{(d)}$ & $E_{\text{CPKMJS}}$ & $E_{\text{TTY}}$ & $E_{\text{GAUSS}}$ & $E_{\text{CPKMJS}}/E_{\text{HFD-HE2}}$ & $E_{\text{GAUSS}}/E_{\text{HFD-HE2}}$ \\ 
 & (Model~IA) & (Model~IB) & (Model~IC) & (Model~II) & (in percent) & (in percent)\\  \hline
 $2$$^{(a)}$  & $-2.645 \times 10^{-9}$ & $-5.147 \times 10^{-9}$ & $-4.183 \times 10^{-9}$ & $-2.6357 \times 10^{-9}$ & 195 & 100 \\
 $3$$^{(b)}$  & $-3.713(3)  \times 10^{-7}$ & $-4.174(5) \times 10^{-7}$ & $-4.006(3) \times 10^{-7}$ & $-3.715(1) \times 10^{-7}$ & 112 & 100 \\
 $4$$^{(c)}$  & $-1.688(1) \times 10^{-6}$ & $-1.815(1) \times 10^{-6}$ &$-1.768(1) \times 10^{-6}$ & $-1.6984(1) \times 10^{-6}$ & 108 & 101 \\
 $5$$^{(c)}$  & $-3.966(1) \times 10^{-6}$ & $-4.201(1) \times 10^{-6}$ & $-4.112(1) \times 10^{-6}$ & $-3.9622(3) \times 10^{-6}$ & 106 & 100 \\
 $6$$^{(c)}$  & $-7.102(2) \times 10^{-6}$ & $-7.467(2) \times 10^{-6}$ & $-7.325(1) \times 10^{-6}$ & $-7.0166(4) \times 10^{-6}$ & 105 & 99 \\
 $7$$^{(c)}$  & $-1.0986(5) \times 10^{-5}$ & $-1.150 (1) \times 10^{-5}$ & $-1.130(1) \times 10^{-5}$ & $-1.0737(1) \times 10^{-5}$ & 106 & 98 \\
 $8$$^{(c)}$  & $-1.5531(6) \times 10^{-5}$ & $-1.621(1) \times 10^{-5}$ & $-1.594(1) \times 10^{-5}$ & $-1.5030(1) \times 10^{-5}$ & 104 & 97 \\
 $9$$^{(c)}$  & $-2.066(1) \times 10^{-5}$ & $-2.152(1) \times 10^{-5}$ & $-2.1176(8) \times 10^{-5}$ & $-1.9830(2) \times 10^{-5}$ & 104 & 96 \\
$10$$^{(c)}$ & $-2.631(1) \times 10^{-5}$ & $-2.736(1) \times 10^{-5}$ & $-2.694(1) \times 10^{-5}$ & $-2.5083(2) \times 10^{-5}$ & 104 & 95 
\end{tabular}
\caption{Ground state energies, in atomic units (columns~2-5), and selected energy ratios, in percent (columns~6-7), at the physical point for various interaction models.
The two-body $s$-wave scattering lengths are
$a_s=234.84$, 
$170.86$,
$188.20$, and
$235.24$~$a_0$ for Model~IA, Model~IB, Model~IC, and Model~II, respectively.
$^{(a)}$The $N=2$ energies are calculated using a grid based approach.
The extrapolated zero imaginary time-step DMC growth energies are
$-2.638(10) \times 10^{-9}$, $-5.11(3) \times 10^{-9}$, $-4.166(10) \times 10^{-9}$, and $-2.632(6) \times 10^{-9}$~a.u. for Model~IA,
Model~IB, Model~IC, and Model~II, respectively.
The comparatively large errors for the $N=2$ DMC energies are due to the large fluctuations associated with extremely weakly-bound systems.
$^{(b)}$For $N=3$,  extrapolated zero imaginary time-step DMC 
growth energies are reported.
$^{(c)}$For $N=4-10$, finite imaginary time-step DMC 
growth energies are reported; the 
errors only account for the statistical uncertainty and not for the 
extrapolation error (the extrapolation 
to the zero imaginary time-step is estimated to lead to a correction that is smaller than
$0.5$~\%). 
$^{(d)}$The HFD-HE2 energies differ slightly from those reported in Ref.~\cite{PhysRevA.102.063320}
due to the difference in $m$~\protect\cite{footnote_mass}. When we use the same mass as Ref.~\cite{PhysRevA.102.063320},
our energies agree within 
errors with those of Ref.~\cite{PhysRevA.102.063320}.}
\label{table1}
\end{table}

\begin{table}
\begin{tabular}{c|ccc| cc}
$N$ & $E_{\text{HFD-HE2}}$$^{(d)}$ & $E_{\text{CPKMJS}}$  & $E_{\text{GAUSS}}$ & $E_{\text{CPKMJS}}/E_{\text{HFD-HE2}}$ & $E_{\text{GAUSS}}/E_{\text{HFD-HE2}}$\\ 
 & (Model~IA) & (Model~IB)  & (Model~II) & (in percent) & (in percent) \\  \hline
 $3$$^{(b)}$  & $-2.656(6) \times 10^{-7}$ & $-2.65(1) \times 10^{-7}$ & $-2.665(1) \times 10^{-7}$      & $100$ & $100$ \\
 $4$$^{(c)}$  & $-1.391(1) \times 10^{-6}$ & $-1.395(5) \times 10^{-6}$ & $-1.4028(1) \times 10^{-6}$   & $100$ & $101$ \\
 $5$$^{(c)}$  & $-3.411(3) \times 10^{-6}$ & $-3.418(4) \times 10^{-6}$ & $-3.4130(2) \times 10^{-6}$   & $100$ & $100$ \\
 $6$$^{(c)}$  & $-6.235(6) \times 10^{-6}$ & $-6.241(4) \times 10^{-6}$ & $-6.1642(4) \times 10^{-6}$   & $100$ & $99$ \\
 $7$$^{(c)}$  & $-9.764(9) \times 10^{-6}$ & $-9.773(4) \times 10^{-6}$ & $-9.5379(6) \times 10^{-6}$   & $100$ & $98$ \\
 $8$$^{(c)}$  & $-1.391(1) \times 10^{-5}$ & $-1.392(8) \times 10^{-5}$ & $-1.3446(1) \times 10^{-5}$   & $100$ & $97$ \\
 $9$$^{(c)}$  & $-1.861(1) \times 10^{-5}$ & $-1.863(10) \times 10^{-5}$ & $-1.7824(1) \times 10^{-5}$ & $100$ & $96$ \\
$10$$^{(c)}$ & $-2.379(2) \times 10^{-5}$ & $-2.382(12) \times 10^{-5}$ & $-2.2626(2) \times 10^{-5}$ & $100$ & $95$ 
\end{tabular}
\caption{Ground state energies, in atomic units
(columns~2-4), and selected energy ratios, in percent (columns~5-6),
 at unitarity for various interaction models.
The superscripts $(b)$, $(c)$, and $(d)$ have the same meaning as in
Table~\protect\ref{table1}.
}
\label{table1_unit}
\end{table}

\end{widetext}

To obtain essentially unbiased 
structural properties, we use the ``forward walking (tagging) scheme'' introduced in Refs.~\cite{Reynolds1986,BARNETT1991258}.
We find that the structural properties calculated in this manner agree, except for regimes where the
sampling probability is extremely low, 
with those obtained by subtracting the VMC estimate $\langle \hat{A} \rangle_{\text{VMC}}$ from twice the 
mixed DMC estimate $\langle \hat{A} \rangle_{\text{DMC}}$~\cite{Reynolds1986}; here, $\langle \hat{A} \rangle_{\text{VMC}}$
and $\langle \hat{A} \rangle_{\text{DMC}}$ denote expectation values of the operator $\hat{A}$
that are calculated with respect to $|\psi_T|^2$ and
$\psi_T \Psi_0$, respectively, where $\Psi_0$ denotes the exact real ground state wave function.

The trial wave function $\psi_T$ is taken to be of the Bijl-Jastrow form
for all four models~\cite{doi:10.1063/1.458943,doi:10.1063/1.460853,doi:10.1063/1.473501},
\begin{eqnarray}
\label{eq_trial}
\psi_T(r_{1,2},\cdots,r_{N-1,N}) = \prod_{j=1}^{N-1} \prod_{k>j}^{N} \exp [ f(r_{j,k})].
\end{eqnarray}
For Model~I, the two-body correlation function $f(r_{j,k})$
contains five variational parameters ($p_{\alpha}$, $p_{\beta}$, $p_{\gamma}$, $p_0$, and $p_1$) that are optimized for each $N$,
\begin{eqnarray}
\label{eq_jastrow_model1}
f_{\text{I}}(r) = -p_{\alpha} r^{-\alpha} - p_{\beta} r^{-\beta} - p_{\gamma} r^{-\gamma}-p_0 \mbox{ln}(r) - p_1 r.
\end{eqnarray}
Two combinations for $\alpha$, $\beta$, and $\gamma$ are considered. The first combination ($\alpha=5$, $\beta=4$, and $\gamma=2$) 
is similar to what has been used frequently in the literature~\cite{doi:10.1063/1.458943,doi:10.1063/1.460853,doi:10.1063/1.473501}, namely, the same 
$\alpha$ and $\gamma$ but $\beta=0$. The second
combination ($\alpha=4.6$, $\beta=1.2$, and $\gamma=0$) 
was found to result in comparable or lower variational energies with one less 
variational parameter.
 %Table~\ref{SI_table2}  
 Table~S1
 in the Supplemental Material
 reports the variational parameters for Model~IB,
 using  $\alpha=4.6$, $\beta=1.2$, and $\gamma=0$ for all $N$.
The VMC energy $E_{\text{VMC}}$ for $N \ge 4$ reaches between $91$~\% and $96$~\% of the DMC energy
 at the physical point and between $91$~\% and $97$~\% of the DMC energy
 at unitarity.

The pair correlation function for the effective low-energy model (Model~II) is known to 
differ from that for the van der Waals potentials. Correspondingly, the functional form of the correlation function needs to be adjusted to capture the short-distance characteristics of the two-body Gaussian potential. 
For Model~II, the two-body correlation function $f(r_{j,k})$
takes the form
\begin{eqnarray}
\label{eq_jastrow_model2}
f_{\text{II}}(r) = 
\bigg\{
\begin{array}{c}
- \sum_{k=3}^8  p_k r^{k-7}  - p_9 \mbox{ln}(r)  \mbox{ for } r>r_m \\
-p_2 r^2 \mbox{ for } r < r_m
\end{array}.
\end{eqnarray}
The parameters $p_3$ and $p_7$ are chosen such that $f_{\text{II}}(r)$ and its first derivative
with respect to $r$ are continuous at $r=r_m$; the matching distance $r_m$
and the parameters  $p_2$, $p_4$, $p_5$, $p_6$, $p_8$, and $p_9$
are optimized for each $N$ by minimizing the energy
(see 
%Table~\ref{SI_table2_gauss} 
Table~S2
in the Supplemental Material).
The VMC energy $E_{\text{VMC}}$ for $N \ge 4$ reaches between  $97$~\% and $98$~\% of the DMC energy
 at both the physical point and 
 at unitarity.
 
For both the realistic van der Waals and low-energy models, we checked carefully 
that the structural properties are independent of the trial wave function. Specifically, we compared 
results for fully optimized and not fully optimized parameters and we compared structural properties obtained by the VMC
method, the mixed estimator, and  a forward walking scheme (see next section).

\subsection{Structural observables}
\label{sec_structural_definition}

This section
defines several structural observables that are analyzed in
Sec.~\ref{sec_results} as a function of $N$ for 
different $V_{\text{int}}$.
As mentioned above, our DMC implementation
determines the structural properties  using a forward walking scheme
that ensures that the excited state contributions contained in the mixed density 
$\psi_T \Psi_0$  decay prior to 
measuring the observable during the DMC run.

The pair distribution function $P_N^{(2)}(r)$ of the $N$-body cluster,
which has units of $(\mbox{length})^{-3}$ and is
normalized according to
\begin{eqnarray}
\int_0^{\infty} P_N^{(2)}(r) r^2 dr =1,
\end{eqnarray}
is obtained by calculating the expectation
value of the operator $\hat{P}_{N}^{(2)}(r)$,
\begin{eqnarray}
\label{eq_pair_operator}
\hat{P}_N^{(2)}(r) = \frac{2}{N(N-1)} \sum_{j=1}^{N-1} \sum_{k>j}^N
\frac{\delta(r_{j,k}-r)}{ r^2}.
\end{eqnarray} 
The short distance behavior of $P_N^{(2)}(r)$
enters into the definition of the $r$-independent scalar two-body contact 
$C_N^{(2)}$~\cite{PhysRevA.101.010501}.
The premise 
is that the short-distance pair correlations of the $N$-body cluster, if scaled by an overall factor,
collapse approximately.
Specifically, the dimensionless two-body contact $C_N^{(2)}$
of the $N$-atom cluster 
is found  by
enforcing~\cite{TAN20082987,TAN20082971,TAN20082952,PhysRevA.101.010501}
\begin{eqnarray}
\label{eq_twobody_contact}
P_N^{(2)}(r) \underset{{\text{small } r}}{\rightarrow} 
C_N^{(2)}  P_2^{(2)}(r). 
\end{eqnarray}
The operator $\hat{P}_N^{(2)}(r)$ defined in Eq.~(\ref{eq_pair_operator})
differs by an overall factor from the operator employed in Ref.~\cite{PhysRevA.101.010501}.
Correspondingly, we convert the results from Ref.~\cite{PhysRevA.101.010501} to our definition
when comparing our two-body contacts with theirs.
Equation~(\ref{eq_twobody_contact}) implies
$C_N^{(2)}=1$ for $N=2$.
In practice, $C_N^{(2)}$ is treated
as a fit
parameter when matching the left and right hand sides of Eq.~(\ref{eq_twobody_contact}), including
only the 
short-distance region
where $P_N^{(2)}(r)$ ($N>2$)
takes values between about 5~\% and 95-100~\% of its maximum. 
 Reference~\cite{PhysRevA.101.010501} extracted the two-body contact for
helium clusters interacting through the realistic LM2M2 potential~\cite{doi:10.1063/1.460139}, an interaction model
that is similar to the Models~IA, IB, and IC used in our work.
Section~\ref{sec_results} determines the two-body contact at the physical point for Models~IA, IB, and IC 
and furthermore discusses that the
two-body contact has limited meaning for $N$-atom clusters interacting through Model~II.
This is not unexpected since $V_{\text{int}}$ for Model~II includes a three-body potential.
    
In addition to the pair distribution function, we monitor the probability 
$\rho^2 P_N^{(\text{jacobi})}(\rho) $ to find one of the particles
located at a distance $\rho$ from the center-of-mass of the other $N-1$ particles.  
The corresponding operator is $\hat{P}_N^{(\text{jacobi})}(\rho)$,
\begin{eqnarray}
\hat{P}_N^{(\text{jacobi})}(\rho) = \frac{1}{N} \sum_{j=1}^{N}
\frac{\delta(\rho_{j}-\rho)}{\rho^2},
\end{eqnarray}
where
\begin{eqnarray}
\rho_j = \left|\vec{r}_j - \frac{1}{N-1} \sum_{k=1,k \ne j}^N \vec{r}_k \right|.
\end{eqnarray}
Since the lowest break-up threshold of the $N$-particle cluster corresponds to the
break-up into a cluster consisting of $N-1$ atoms and a single far-separated atom,  $P_N^{(\text{jacobi})}(\rho)$
should---in the large $\rho$ limit---fall off as
\begin{eqnarray}
\label{eq_an_const}
P_N^{(\text{jacobi})}(\rho)
\underset{{\text{large } \rho}}{\rightarrow}
A_N \rho^{-2} \exp(-2 \kappa_N \rho),
\end{eqnarray}
where the binding momentum $\kappa_N$ is defined through 
$\sqrt{2 \mu_N \epsilon_N}/\hbar$,
the 
binding energy $\epsilon_N$ of the $N$-particle cluster is defined with respect to the
ground state energy $E_{N-1}$ of the $N-1$ cluster, and $\mu_N$
is equal to $(N-1)	m/N$.
By comparing the tail of $P_N^{(\text{jacobi})}(\rho)$
with the expected asymptotic behavior, the extent of the universal, 
binding-energy-dominated regime can be determined.
We note that 
$\lim_{r \rightarrow \infty} P_N^{(2)}(r)$ and $\lim_{\rho \rightarrow \infty} P_N^{(\text{jacobi})}(\rho)$ 
behave, except for an overall
normalization constant, identically.
For ground state helium clusters with $N \ge 3$, the $r$-region over which 
$P_N^{(2)}(r)$ is governed by the binding momentum is notably smaller than
the $\rho$-region over which $P_N^{(\text{jacobi})}(\rho)$ is governed by the binding momentum.
This can be seen by rewriting $\rho_j$,
\begin{eqnarray}
\rho_j = 
\left| 
\frac{1}{N-1}
\sum_{k=1,k \ne j}^N \vec{r}_{j,k} \right|.
\end{eqnarray}
For $\rho_j \rightarrow \infty$, the vectors $\vec{r}_{j,k}$ are all parallel and 
$P_N^{(\text{jacobi})}(\rho)$ and $P_N^{(2)}(r)$
agree, except for an overall normalization factor.
When $\rho_j$ is finite,
the vectors $\vec{r}_{j,k}$ with $k=1,\cdots,j-1,j+1,\cdots,N$ are not all parallel and
 $P_N^{(2)}(r)$ deviates from $P_N^{(\text{jacobi})}(\rho)$.

To quantify the three-body correlations of the $^4$He$_N$ clusters, we monitor 
two complementary distribution functions, $P_N^{(3,\text{jacobi})}(\rho_3)$
and $P_N^{(3,\text{shape})}(\bar{x},\bar{y})$. 
The three-body Jacobi distribution function $P_N^{(3,\text{jacobi})}(\rho_3)$, which  
is measured by the 
operator $\hat{P}_N^{(3,\text{jacobi})}(\rho_3)$,
\begin{eqnarray}
\hat{P}_N^{(3,\text{jacobi})}(\rho_3) = 
\frac{2}{N(N-1)(N-2)}
\sum_{j=1}^{N-2}\sum_{k>j}^{N-1} \sum_{l>k}^N \nonumber \\
\left[ \frac{\delta(\rho_{jk,l}-\rho_3)  }{ (\rho_3)^2} + 
\frac{\delta(\rho_{jl,k}-\rho_3)  }{ (\rho_3)^2} + 
\frac{\delta(\rho_{kl,j}-\rho_3)  }{ (\rho_3)^2}
\right]
,
\end{eqnarray}
where
\begin{eqnarray}
\rho_{jk,l} = \left | \vec{r}_l - \frac{1}{2} \left(
\vec{r}_j+\vec{r}_k 
\right) \right | .
\end{eqnarray} 
The quantity $(\rho_3)^2 P_N^{(3,\text{jacobi})}(\rho_3) $
tells us, for each triple within the $N$-body cluster, 
the likelihood to find one of the particles at distance $\rho_3$ from the center of mass of
the other two particles of the triple.
In analogy to the two-body contact $C_N^{(2)}$, we define a 
$(2+1)$ or pair-atom contact $C_N^{(2+1)}$ for $N \ge 3$ through 
\begin{eqnarray}
\label{eq_2plus1_contact}
P_N^{(3,\text{jacobi})} (\rho_3) 
\underset{{\text{small } \rho}_3}{\rightarrow}
C_N^{(2+1)} P_3^{(3,\text{jacobi})}.
\end{eqnarray}

Equation~(\ref{eq_2plus1_contact}) defines the pair-atom contact 
$C_N^{(2+1)}$ through the
short-range behavior of the distribution function.
Alternatively, we may define $C_N^{(2+1)}$
by assuming that the many-body wave function
$\Psi$  factorizes when
$\rho_{jk,l}$ takes on small values,
\begin{eqnarray}
\label{eq_product_ansatz}
\Psi(\vec{r}_1,\cdots,\vec{r}_N)
\underset{{\text{small } \rho}_{jk,l}}{\rightarrow} \nonumber \\
\Phi(\vec{\rho}_{jk,l}) B_N^{(2+1)}(\vec{r}_{j,k}, \vec{R}_{j,k,l} ,\left\{ \vec{r}_{n;n\ne j,k,l} \right\}),
\end{eqnarray}
where $\vec{R}_{j,k,l}=(\vec{r}_j+\vec{r}_k+\vec{r}_l)/3$.
The function $B_N^{(2+1)}$ is non-universal and the limit
in Eq.~(\ref{eq_product_ansatz}) is taken while keeping $\vec{r}_{j,k}$, $\vec{R}_{j,k,l}$,
and all $\left\{ \vec{r}_{n;n\ne j,k,l} \right\}$ unchanged.
If the pair-atom function $\Phi(\vec{\rho}_{jk,l})$ is universal, then 
the pair-atom contact is a meaningful quantity and can be related to $\Phi(\vec{\rho}_{jk,l})$
following the same steps as when relating the two-body contact, 
the relevant product ansatz, and the pair distribution function (see, e.g., Ref.~\cite{PhysRevA.101.010501}). 
While the $(2+1)$ contact characterizes three-body correlations of $N$-particle systems, it differs conceptually from the three-body contact
considered in the literature~\cite{PhysRevA.86.053633,PhysRevLett.106.153005}.

Since the distribution function $P_N^{(3,\text{jacobi})}(\rho_3) $ does not capture
the relative orientation of the sub-trimers
(the angles are being averaged over), we additionally monitor the 
normalized trimer correlation function
$P_N^{(3,\text{shape})}(\bar{x},\bar{y})$, which captures the 
relative orientation of any three atoms within the $N$-atom cluster~\cite{doi:10.1126/science.aaa5601}.
For each triple spanned by $\vec{r}_j$, $\vec{r}_k$, and $\vec{r}_l$,
we determine the maximum of $r_{j,k}$, $r_{j,l}$, and $r_{k,l}$
and scale all lengths by this value. For concreteness, let us assume that  $r_{j,k}$ is larger than $r_{j,l}$ and $r_{k,l}$.
Next, we rotate the triangle spanned by 
 $\vec{r}_j$, $\vec{r}_k$, and $\vec{r}_l$ so that it lies in the $xy$-plane, so
 that the normalized position vectors
of particles $j$ and $k$ are equal to $(x,y,z)=(\pm 1/2,0,0)$, and
so that the $y$-coordinate of
particle $l$ is positive. The distribution
$P_N^{(3,\text{shape})}(\bar{x},\bar{y})$ yields the likelihood that the $l$th particle
has the normalized, rotated  position vector $(\bar{x},\bar{y},0)$.

The hyperradial distribution function $P_N^{(\text{hyper})}(\rho_N)$
is measured by the
operator $\hat{P}_N^{(\text{hyper})}(\rho_N)$,
\begin{eqnarray}
\hat{P}_N^{(\text{hyper})}(\rho_N) = \frac{\delta(\rho_N-R)}{ R^{3N-4}},
\end{eqnarray} 
where $R$ is the hyperradius,
\begin{eqnarray}
\label{eq_hyperradius}
R^2 = \frac{1}{N^2} \sum_{j=1}^{N-1} \sum_{k>j}^N r_{j,k}^2.
\end{eqnarray}
The
normalization is such that
\begin{eqnarray}
\int_0^{\infty} P_N^{(\text{hyper})}(\rho_N) (\rho_N)^{3N-4} d \rho_N = 1.
\end{eqnarray}
The quantity $(\rho_N)^{3N-4} P_N^{(\text{hyper})}(\rho_N) $
 tells one the likelihood that the $N$-atom
cluster has the hyperradius $\rho_N$.
The hyperradius provides a measure of the cluster size~\cite{LIN19951,doi:10.1063/1.481404,D_Incao_2018}.
Our definition of the hyperradius implies a hyperradial mass of $M$,
\begin{eqnarray}
M = N m.
\end{eqnarray}
Section~\ref{sec_results} uses the hyperradial distribution functions to determine approximate effective
hyperradial potential curves assuming separability of the hyerradial
and hyperangular degrees of freedom.
Despite the crudeness of the approach (the coupling of the hyperradial and hyperangular degrees of freedom can, in general,
not be neglected), the resulting
approximate hyperradial potential curves provide, as shown in Sec.~\ref{sec_results}, 
some insight.

The kinetic energy release (KER),
\begin{eqnarray}
\text{KER} = \sum_{j=1}^{N-1} \sum_{k>j}^{N} \frac{1}{r_{j,k}},
\end{eqnarray}
of the helium dimer as well as pure and mixed-isotope helium trimers has been measured in 
Coulomb explosion experiments~\cite{Voigtsberger2014,doi:10.1126/science.aaa5601,Zeller14651}. 
While it is not 
clear that the 
experimental determination of the KER generalizes 
straightforwardly to larger clusters~\cite{Ulrich2011,PhysRevA.98.050701},
Sec.~\ref{sec_results} reports and interprets the KER for helium clusters with up to $N=10$ particles.

\section{Results}
\label{sec_results}

This section presents results for the observables defined in Sec.~\ref{sec_structural_definition}.
In addition to tracking the structural properties
as a function of $N$, particular focus is placed on comparing
\begin{itemize}
\item the characteristics of helium clusters at the physical point 
(``true''  helium clusters) and quantum clusters at unitarity 
(helium-helium interaction artificially tuned to unitarity);
\item the 
characteristics of helium clusters at the physical point  interacting through 
the three realistic interaction potentials Model~IA, Model~IB, and Model~IC; 
\item the characteristics of helium clusters at the physical point interacting through 
the realistic HFD-HE2 potential (Model IA) and the effective low-energy potential
(Model II); and
\item the characteristics of helium clusters at unitarity interacting through 
the realistic HFD-HE2 potential 
and the effective low-energy potential.
\end{itemize}

To put the structural properties into context,
we discuss a few characteristics of the energies 
at the physical point (see Table~\ref{table1}) and
 at
unitarity (see Table~\ref{table1_unit}).
Table~\ref{table1} shows that
the two-body binding energy for the CPKMJS potential at the physical point is $1.95$ times larger than that
for the HFD-HE2 potential.
For $N=3$, the difference in the energy is notably smaller, namely the energy
for the CPKMJS potential at the physical point is $12$~\% larger than that
for the HFD-HE2 potential.
As $N$ increases, the difference decreases from $8$~\% for $N=4$ to
$4$~\% for $N=10$. For $N=10$, this percentage difference between the energy for Model~IB and Model IA
is similar to that between the energy for the
effective low-energy Model~II and Model~IA.
Table~\ref{table1_unit} shows that 
the dependence of the energy at unitarity is, for the scaled realistic interaction potentials, 
notably suppressed
compared to the physical point.
Specifically, the energies for $N=3-10$ for the CPKMJS potential are 
slightly larger than those  for the HFD-HE2 potential (rounding, the percentage is $100$~\%).

\begin{widetext}

\begin{figure}
\includegraphics[width=0.9\textwidth]{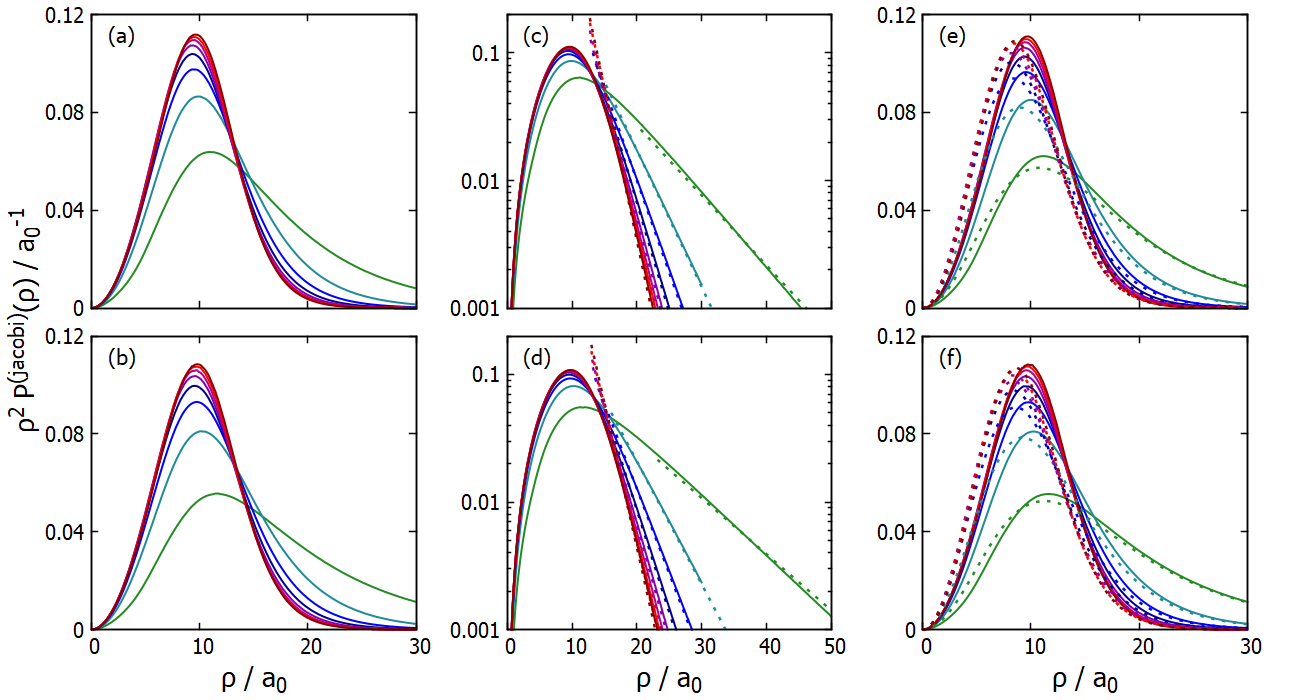}
\caption{$\rho^2 P_N^{(\text{jacobi})}(\rho)$  for $N=3-10$ at the physical point (top
row) and
at unitarity (bottom row).
The solid lines in Figs.~\protect\ref{fig_jacobi}(a)-\protect\ref{fig_jacobi}(d) 
show $\rho^2 P_N^{(\text{jacobi})}(\rho)$ for Model~IB. At 
$\rho = 20$~$a_0$, the curves are ordered, from top to bottom, from the 
smallest $N$ (the curve for $N=3$ is green) to the largest $N$ (the curve for
$N=10$ is dark red). 
Note that the data shown in the first and second columns are identical but
that the $x$- and $y$-scales differ.
The dotted lines in Figs.~\protect\ref{fig_jacobi}(b)
and \protect\ref{fig_jacobi}(d) show the asymptotic behavior 
$A_N \exp(-2 \kappa_N \rho)$, using the numerically
determined
binding momentum $\kappa_N$ and treating the ``normalization constant'' 
$A_N$ as a fitting parameter.
The solid and dotted lines in Figs.~\protect\ref{fig_jacobi}(e)-\protect\ref{fig_jacobi}(f) (third 
column) show $\rho^2 P_N^{(\text{jacobi})}(\rho)$ for Model~IA and Model~II, 
respectively. The agreement between the dotted and solid lines is 
excellent at large $\rho$ and deteriorates for smaller $\rho$. The deterioration
is due to the
inability of the low-energy model to fully capture the small length scale
correlations.
The color scheme used here is also used in Figs.~\protect\ref{fig_pair_all},
\protect\ref{fig_pair_short}, \protect\ref{fig_jacobi3_alt}(a), \protect\ref{fig_hyper}, 
\protect\ref{fig_ker},  
%\protect\ref{fig_jacobi3} 
S3, and S4.
The layout used here, i.e.,
the top row showing results at the physical point and the bottom row
showing results at unitarity, is also used in
Figs.~\protect\ref{fig_pair_all},
\protect\ref{fig_shape}, \protect\ref{fig_hyper}, \protect\ref{fig_ker}, and 
%\protect\ref{fig_jacobi3}.
S4.
}
\label{fig_jacobi}
\end{figure}    

\end{widetext}

The solid lines in Fig.~\ref{fig_jacobi} show the likelihood 
$\rho^2 P_{N}^{(\text{jacobi})}(\rho)$ for realistic interaction models to find a particle at
distance $\rho$ from the center of mass of the other $N-1$ particles
for $N=3-10$.
The color of the lines changes nearly continuously from green for $N=3$ to dark red
for $N=10$.
The top and bottom rows show results at the physical point and at unitarity,
respectively. It can be seen that the distributions at unitarity extend to somewhat larger $\rho$,
owing to the smaller binding energies $\epsilon_N$ at unitarity than at
the physical point.
The third column compares results for the HFD-HE2 potential and the
effective low-energy model.
It can be seen that the large $\rho$ behavior
of $\rho^2 P_{N}^{(\text{jacobi})}(\rho)$ for the HFD-HE2 potential (Model~IA, solid lines) and
for the
effective low-energy potential (Model~II, dotted lines) agrees well.
This is expected since the effective low-energy potential has been shown to
reproduce the energies of the $N$-particle cluster interacting through the
HFD-HE2 potential at the 95~\% or higher level (see Ref.~\cite{PhysRevA.102.063320}
and Tables~\ref{table1} and \ref{table1_unit}).

The solid lines in the left and middle columns 
of Fig.~\ref{fig_jacobi} show $\rho^2 P_{N}^{(\text{jacobi})}(\rho)$
for the realistic CPKMJS potential (Model~IB) on a linear and logarithmic scale,
respectively. 
The logarithmic representation allows us to visually quantify the portion of the distribution
that is governed by the exponential binding momentum dominated fall-off.
Specifically, the dotted lines show the expected fall-off, using the 
binding momentum
$\kappa_N$, obtained by combining DMC energies of clusters containing 
$N$ and $N-1$ atoms, as input. To plot the dotted lines, the normalization constant
$A_N$, Eq.~(\ref{eq_an_const}), is  treated as a fitting parameter to best match the large-$\rho$
tail, including $\rho \ge \rho_{m}$, where
$\rho_{m}$ is adjusted such that
$\int_{\rho_{m}}^{\infty} P_{N}^{(\text{jacobi})}(\rho)  \rho^2 d \rho$ is equal to $0.2$.
The visual agreement at large $\rho$ between the solid and dotted lines in the middle column of
Fig.~\ref{fig_jacobi} is good.

\begin{widetext}

\begin{figure}
\includegraphics[width=0.9\textwidth]{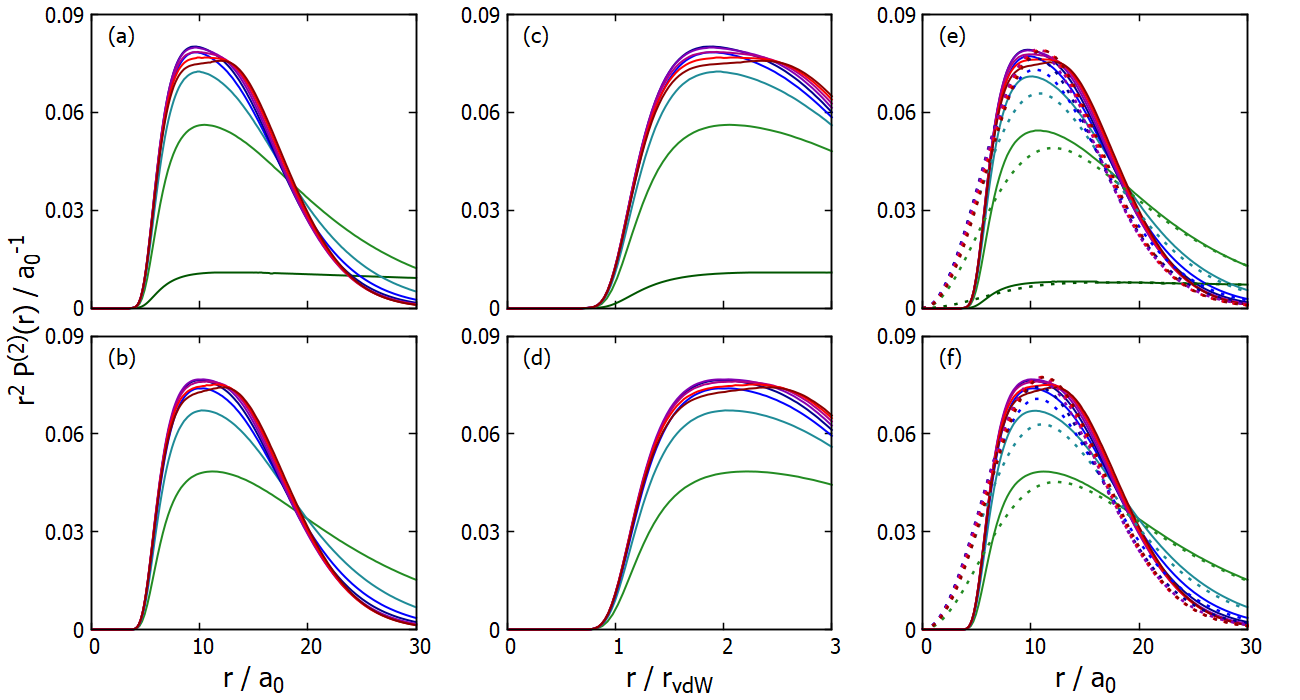}
\caption{$r^2 P_N^{(2)}(r)$  for $N=2-10$ at the physical point (top
row) and  for $N=3-10$
at unitarity (bottom row).
The solid lines in Figs.~\protect\ref{fig_pair_all}(a)-\protect\ref{fig_pair_all}(d) 
show $r^2 P_N^{(2)}(r)$ for Model~IB. 
Note that the data shown in the first and second columns are identical: the
first column 
shows $r$ in units of $a_0$ and the second column shows $r$ in units of $r_{\text{vdW}}$,
focusing on the small-$r$ region.
It can be seen that the scaled pair distribution functions for different $N$
collapse approximately for $r \approx r_{\text{vdW}}$. 
The solid and dotted lines in Figs.~\protect\ref{fig_jacobi}(e)-\protect\ref{fig_jacobi}(f) (third 
column) show $r^2 P_N^{(2)}(r)$ for Model~IA and Model~II, 
respectively. 
Differences are most pronounced in the $r \lesssim 20$~$a_0$ region.
The color scheme is the same as in Fig.~\ref{fig_jacobi}.
}
\label{fig_pair_all}
\end{figure}    

\end{widetext}

Figure~\ref{fig_pair_all} shows the scaled
pair distribution functions $r^2 P_N^{(2)}(r)$ for $N=2-10$ at the physical point
(top row) and at unitarity (bottom row).
The scaled pair distribution functions for the realistic interaction models
display a clear maximum for $N \lesssim 6$. For larger $N$, the maximum broadens and
shifts to larger $r$-values; for these larger $N$, the scaled pair distribution
functions display a hint of a double-peak structure that can be interpreted as a signature of the 
development of a ``second length scale or shell''. It is important to keep in mind that the 
clusters at the physical point and at unitarity 
are extremely floppy and diffuse and that the terms ``second length scale'' and
``second shell'' should be contextualized within the framework of extremely diffuse quantum liquids.
The double-peak structure is not reproduced by the low-energy model (dotted lines in the third column).

The third column of Fig.~\ref{fig_pair_all} shows that the quantities $r^2 P_N^{(2)}(r)$
for Model~IA (solid lines) and Model~II (dotted lines) differ
for small $r$ ($r \lesssim 20$~$a_0$). Interestingly, the 
scaled pair distribution functions for the HFD-HE2 potential and the CPKMJS potential (solid lines) 
rise at about the same $r$-value for all $N$, namely at $r \approx 4.5$~$a_0$
or $r \approx 0.9$~$r_{\text{vdW}}$.  
Careful inspection shows that the rise is shifted to somewhat larger $r$-values for the clusters 
at unitarity
interacting through realistic potentials  than for the clusters at the physical point  interacting
through realistic potentials.
The scaled
pair distribution functions for the effective low-energy potential (Model~II, dotted lines), 
in contrast, rise at much smaller $r$ values.
The scaled pair distribution functions
for Model~I and Model~II are different at small $r$ for two reasons:
(i) The two-body Gaussian potential used in Model~II does not have a hard wall at small $r$.
(ii)
Model~II contains a repulsive three-body Gaussian potential, which alters the behavior when three 
particles are in close vicinity to each other, impacting the short-distance correlations of two-, three-, and higher-body subclusters.

\begin{figure}
\includegraphics[width=0.4\textwidth]{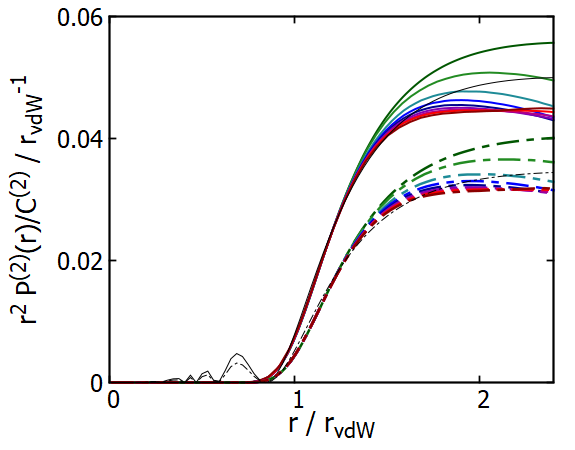}
\caption{$ r^2 P_N^{(2)}(r)/C_N^{(2)}$, $N=2-10$,  
for two realistic interaction potentials at the physical point.
The sets of dash-dotted and  solid lines show
results for Model~IA and Model~IB, respectively.
The color scheme is the same as in Fig.~\ref{fig_jacobi}.
The thin black dash-dotted and solid lines  show the
universal van der Waals function
$|\varphi_{\text{vdW}}(r)|^2$, Eq.~(\ref{eq_vdw}),
for Model~IA and Model~IB, respectively (these two models are characterized by slightly
different $r_{\text{vdW}}$); 
the normalization constant $B$ 
is adjusted by fitting  $|\varphi_{\text{vdW}}(r)|^2$ to $r^2 P_2^{(2)}(r)$,
including $r$ values ($r \lesssim 2 r_{\text{vdW}}$) for which  $P_2^{(2)}(r)$
takes values that are larger than 5~\% and smaller than the maximum of  $P_2^{(2)}(r)$
for Model~IA and smaller than 95~\%
of the maximum of  $P_2^{(2)}(r)$ for Model~IB, respectively.
}
\label{fig_pair_short}
\end{figure}

To highlight the universality of the short-range behavior of the scaled pair distribution function
$r^2 P_N^{(2)}(r)$
for realistic interaction models, 
Fig.~\ref{fig_pair_short} replots 
$r^2 P_N^{(2)}(r)$ at the physical point---including the factor  $C_N^{(2)}$---for
Model~IA (dash-dotted lines) and Model~IB (solid lines).
As discussed in Sec.~\ref{sec_theory},
the two-body contact $C_N^{(2)}$ is determined by fitting the $N>2$ curves 
for small $r$ to the $N=2$ curve.
It can be seen that the rise of the scaled curves collapses for $N=2-10$ in the regime
$r \lesssim 1.4$~$ r_{\text{vdW}}$ separately for both interaction models.
The fact that the curves  for each of the interaction models collapse
confirms that the two-body contact $C_N^{(2)}$, determined in the manner
described in Sec.~\ref{sec_theory}, provides a meaningful characterization of the
short-distance behavior of van der Waals clusters.

%Table~\ref{SI_table_contact}
Table~S3
reports $C_N^{(2)}$ for helium clusters with $N=3-10$ interacting
through Model~IA-IC at the physical point.
The ratio of $C_N^{(2)}$, $N \ge 3$, for 
two different interaction potentials is approximately constant. To leading order, this 
ratio is given by the ratio of $a_s/r_{\text{vdW}}$ for the two different interaction potentials.
Specifically, the values for the HFD-HE2 potential are between $1.32$ and $1.38$ 
times larger than those for the CPKMJS potential; for comparison, 
$(a_s/r_{\text{vdW}})_{\text{HFD-HE2}}/(a_s/r_{\text{vdW}})_{\text{CPKMJS}}$ is equal to $1.40$. Those for 
the TTY potential are between $1.09$ and $1.10$ times larger than those for the CPKMJS potential;
for comparison, 
$(a_s/r_{\text{vdW}})_{\text{TTY}}/(a_s/r_{\text{vdW}})_{\text{CPKMJS}}$ is equal to $1.10$.

To understand this behavior, we recall that the pair distribution functions for the realistic interaction potentials 
at the physical point are, for
$N \gtrsim 5$, to a very good approximation independent of the potential model
[compare, e.g., the solid lines in
Figs.~\ref{fig_pair_all}(a) and \ref{fig_pair_all}(e)]. The $N=2$ pair distribution functions, in contrast,
differ notably. Because of this, the difference between the contacts $C_{N}^{(2)}$, $N \gtrsim 5$,  for Model~IA and Model~IB
predominantly reflects the difference between the respective $N=2$ pair distribution functions. Specifically,
using the fact that the dimers are weakly bound and the pair distribution functions are normalized, the difference
in the height of $r^2 P_2^{(2)}(r)$ at small $r$ for different realistic potential models can be expressed in terms of the binding momentum
and thus, using effective range theory, in terms of $a_s/r_{\text{vdW}}$. 
Assuming that the pair distribution functions for different potential models agree for larger $N$, we find 
that the ratio of the two-body contacts for larger $N$ is given, to leading order, by the ratio between $a_s/r_{\text{vdW}}$ for the two interaction potentials.
Our analysis indicates that the $N$-dependence of the two-body contact $C_N^{(2)}$ 
for helium clusters at the physical point interacting through one realistic interaction model is,
to a fairly good approximation, universally linked to that for 
helium clusters interacting through another realistic interaction model.
The arguments presented here are reminiscent of the discussion
of effective range corrections to the asymptotic normalization   
constant, which is defined by relating the ``true'' nuclear  wave function to a wave function
that is calculated assuming that the 
effective interaction in the asymptotically dominant channel has vanishing 
range~\cite{KimTubis,PhysRevC.25.1616}.

%Table~\ref{SI_table_contact}
Table~S3
also compares our results  with those
obtained in Ref.~\cite{PhysRevA.101.010501} for the LM2M2 potential.
The $C_N^{(2)}$ values for the LM2M2  potential are between $1.06$ and $1.08$ 
times larger than those for the CPKMJS potential; this is quite a bit smaller than
$(a_s/r_{\text{vdW}})_{\text{LM2M2}}/(a_s/r_{\text{vdW}})_{\text{CPKMJS}}=1.13$.
We expect that the LM2M2 data from Ref.~\cite{PhysRevA.101.010501}
would follow the same trends as displayed by our data; we speculate that the 
differences might be related to the different data analysis strategies employed.

Last, we note that our
analysis of the short-distance behavior of the scaled pair distribution functions for
the effective low-energy
Model~II reveals that the small-$r$ behaviors of $r^2 P_2^{(2)}(r)$
and $r^2 P_N^{(2)}(r)$ with $N \ge 3$ do not collapse as neatly by introducing an
$r$-independent scaling factor for each $N$
(see 
%Fig.~\ref{SI_fig_pair_gauss}) 
Fig.~S3 from the Supplemental Material)
as the corresponding data for the realistic interaction models. Due to the presence of the repulsive three-body potential, the low-energy model does not capture the 
small-$r$, ``high-energy'' van der Waals universality of the pair distribution function.

The fact 
that
the curves for Model~IA in Fig.~\ref{fig_pair_short} are pushed to larger $r$ compared to those for Model~IB
can be interpreted as being due to
Model~IA being  characterized by
a larger effective repulsion  than Model~IB:
the two-body $s$-wave scattering length  for Model~IA 
is larger than that for Model~IB ($a_s= 234.84$~$a_0$ compared to $a_s=170.86$~$a_0$).
Interestingly, the rise of the scaled pair distribution functions
is captured quantitatively by the universal van der Waals function
$\varphi_{\text{vdW}}(r)$~\cite{PhysRevA.59.1998,PhysRevA.58.4222},
\begin{eqnarray}
\label{eq_vdw}
\varphi_{\text{vdW}}(r) =
B \bigg[ 
\Gamma(5/4) x^{1/2} J_{1/4}(2 x^{-2}) -  \nonumber \\
\frac{r_{\text{vdW}}}{a_s} \Gamma(3/4) x^{1/2} J_{-1/4}( 2 x^{-2})
\bigg],
\end{eqnarray}
which is obtained by solving the scaled radial Schr\"odinger equation for a 
purely attractive $-C_6/r^6$ potential. 
In Eq.~(\ref{eq_vdw}), $x$ is equal to $r/r_{\text{vdW}}$. 
Thin black dash-dotted and solid lines 
in Fig.~\ref{fig_pair_short} show the quantity
$|\varphi_{\text{vdW}}(r) |^2$   for Model~IA 
($a_s/r_{\text{vdW}} = 46.95$) and Model~IB ($a_s/r_{\text{vdW}} = 33.63$), respectively.
The nodes of the
wave function $\varphi_{\text{vdW}}(r)$ 
in the $ r \lesssim r_{\text{vdW}}$ region reflect the presence of deep-lying 
two-body bound states.
For $r$-values beyond the last node,
the density $|\varphi_{\text{vdW}}(r)|^2$ 
agrees well
with $ r^2 P_2^{(2)}(r)/C_N^{(2)}$.
For the infinite
scattering length case, Refs.~\cite{PhysRevLett.112.105301,PhysRevA.90.022106}
established the van der Waals universality of the
short-distance correlations of the scaled pair distribution function of trimers
interacting through realistic interaction potentials.
Figure~\ref{fig_pair_short} shows that 
$|\varphi_{\text{vdW}}(r)|^2$
captures the short-distance correlations of $r^2 P_N^{(2)}(r)$
also for helium clusters at the physical point.

Figures~\ref{fig_jacobi3_alt} and \ref{fig_shape} as well as 
%Fig.~\ref{fig_jacobi3} 
Fig.~S4
in the Supplemental Material
summarize the three-body correlations of $N$-atom clusters.
%Figure~\ref{fig_jacobi3}, 
Figure~S4, 
which shows the 
quantity $(\rho_3)^2 P_N^{(3,\text{jacobi})}(\rho_3)$, 
highlights two key points.
First, the scaled three-body distributions $(\rho_3)^2 P_N^{(3,\text{jacobi})}(\rho_3)$ 
for Model~IB and Model~II (first and third columns) are visually indistinguishable, including in the small $\rho_3$ region; this is 
in clear contrast to the behavior of the scaled pair distribution functions.
Second, the quantity $(\rho_3)^2 P_N^{(3,\text{jacobi})}(\rho_3)$ becomes narrower as $N$ 
changes from $N=3$ to $N=4$ to $N=5$ but changes comparatively little for $N=6-10$.
This indicates that the correlations of the three-body sub-system saturate approximately for these $N$-values.
This ``saturation'' is 
different from the behavior of the scaled pair distribution functions,
which show a more pronounced $N$ dependence for $N=6-10$.

Figure~\ref{fig_jacobi3_alt}(a) focuses on the small $\rho_3$ behavior at the physical
point. The solid and dash-dotted lines show 
$[(\rho_3)^2 P_N^{(3,\text{jacobi})}(\rho_3)/C_N^{(2+1)}]/r_{\text{vdW}}^{-1}$ for $N=3-10$ for
Model~IA and Model~II,
respectively; to make the figure, the $x$- and $y$-axis are scaled using the
van der Waals length $r_{\text{vdW}}$ for the HFD-HE2 potential (Model~IA).
The collapse of the scaled distribution functions is extremely clean for the realistic
interaction potential (solid lines) and very clean for the low-energy potential (dash-dotted lines).
Differences between the scaled curves for the realistic and low-energy models are clearly
visible for small $\rho_3$.
Figures~\ref{fig_jacobi3_alt}(b) and \ref{fig_jacobi3_alt}(c)
show the $N$-dependence of the $(2+1)$ contact $C_N^{(2+1)}$ at the physical point and at unitarity, respectively,
for three different interaction potentials. The overall trends are the same for all three interaction potentials:
$C_N^{(2+1)}$ increases for $N \lesssim 6$ or $7$ and then slowly decreases. 
The contacts $C_N^{(2+1)}$ for the low-energy model (triangles) are notably larger for $N\ge 4$ than those for
the realistic potentials (squares and circles). Interestingly, while the contacts $C_N^{(2+1)}$
for the two realistic potentials (Model~IA and Model~IB) differ by a small amount for
$N \ge 4$ at the physical point, they coincide, within our numerical accuracy, at unitarity
(see numerical values of $C_N^{(2+1)}$ are collected in 
%Table~\ref{SI_table_contact_2plus1}). 
Table~S4
This behavior of the contact is related to the three-body energies. The ratio
$E_{\text{CPKMJS}}/E_{\text{HFD-HE2}}$ is equal to $1.12$ at the physical point
(see Table~\ref{table1}) 
and $1.00$ at unitarity (see Table~\ref{table1_unit}).

\begin{figure}
\includegraphics[width=0.4\textwidth]{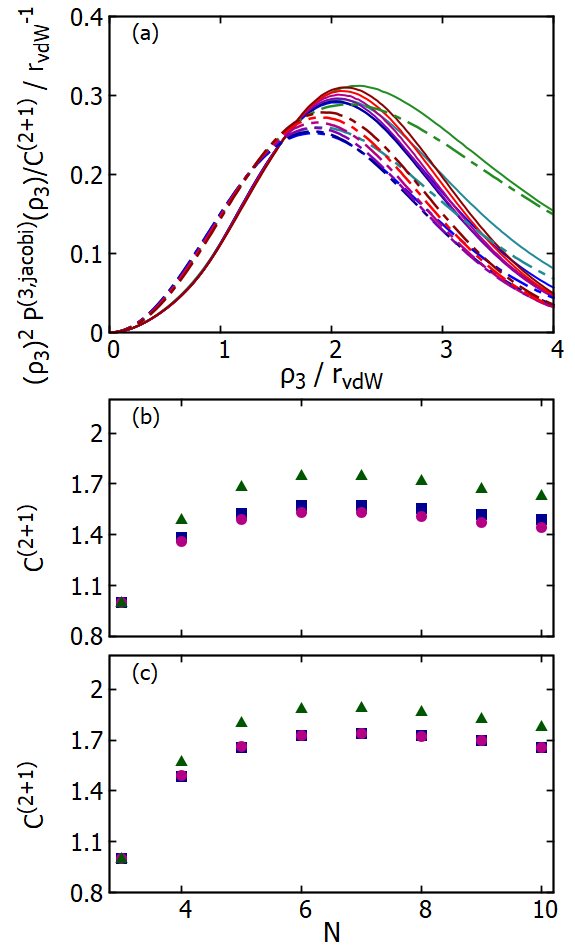}
\caption{Triple correlations and $(2+1)$ contact.
(a) The solid and dash-dotted lines show
$(\rho_3)^2 P_N^{(3,\text{jacobi})}(\rho_3) / C_N^{(2+1)}$   
for the realistic HFD-HE2 potential (Model~IA) and the effective low-energy potential
(Model~II), respectively, 
at the physical point for $N=3-10$.
The color scheme is the same as in Fig.~\ref{fig_jacobi}.
(b) The squares, circles, and triangles show the pair contact $C_N^{(2+1)}$ at the physical point
as a function of $N$ for Model~IA, Model~IB, and Model~II, respectively.
(c) The squares, circles, and triangles show the pair contact $C_N^{(2+1)}$ at unitarity
as a function of $N$ for Model~IA, Model~IB, and Model~II, respectively.
}
\label{fig_jacobi3_alt}
\end{figure}    

As already mentioned in Sec.~\ref{sec_theory},
the $(2+1)$ contact investigated here differs from the three-body contact
investigated in Ref.~\cite{PhysRevA.86.053633,PhysRevLett.106.153005} at the physical point.
While the three-body contact for realistic
interaction models is, to a large degree, governed by the short-distance two-body correlations,
the three-body contact for the low-energy model depends notably on the repulsive three-body potential.
The $(2+1)$ contact, in contrast, captures the behavior as a third particle approaches the center-of-mass of
a two-body sub-unit of any size. As such, the $(2+1)$ contact probes, on average, larger length scales than the
three-body contact. Correspondingly, the low-energy model does a better job of 
reproducing the $(2+1)$ contact obtained for the realistic potentials than it does of reproducing 
the three-body contact 
obtained for the realistic potentials
(we are not showing data for the three-body contact).

To gain insights into the distribution of the shapes that the triples are arranged
in, the first, second, and third columns of
Fig.~\ref{fig_shape} show the distribution
function $P_N^{(3,\text{shape})}(\bar{x},\bar{y})$ for $N=3$, $N=4$, and $N=10$, respectively.
We observe that the distributions, and thus the structures, at the physical point (top row)
and at unitarity (bottom row) are very similar. The highest probability is found at 
$\bar{x}=0$ and $\bar{y} \approx 0.35$, which corresponds to a slightly
elongated triangle. Even though the distributions have a maximum, the clusters' wave functions
include essentially all shapes, except for those where two particles sit on top of each other
($\bar{y}=0$ and arbitrary $\bar{x}$) and where the triangles are highly elongated 
($\bar{y} \approx 0$ and $\bar{x} \approx \pm 0.5$). 
Figure~\ref{fig_shape} shows that the distributions become more peaked 
with
increasing $N$
and that the likelihood to find highly-elongated triangles becomes smaller with
increasing $N$.

\begin{widetext}

\begin{figure}
\includegraphics[width=0.9\textwidth]{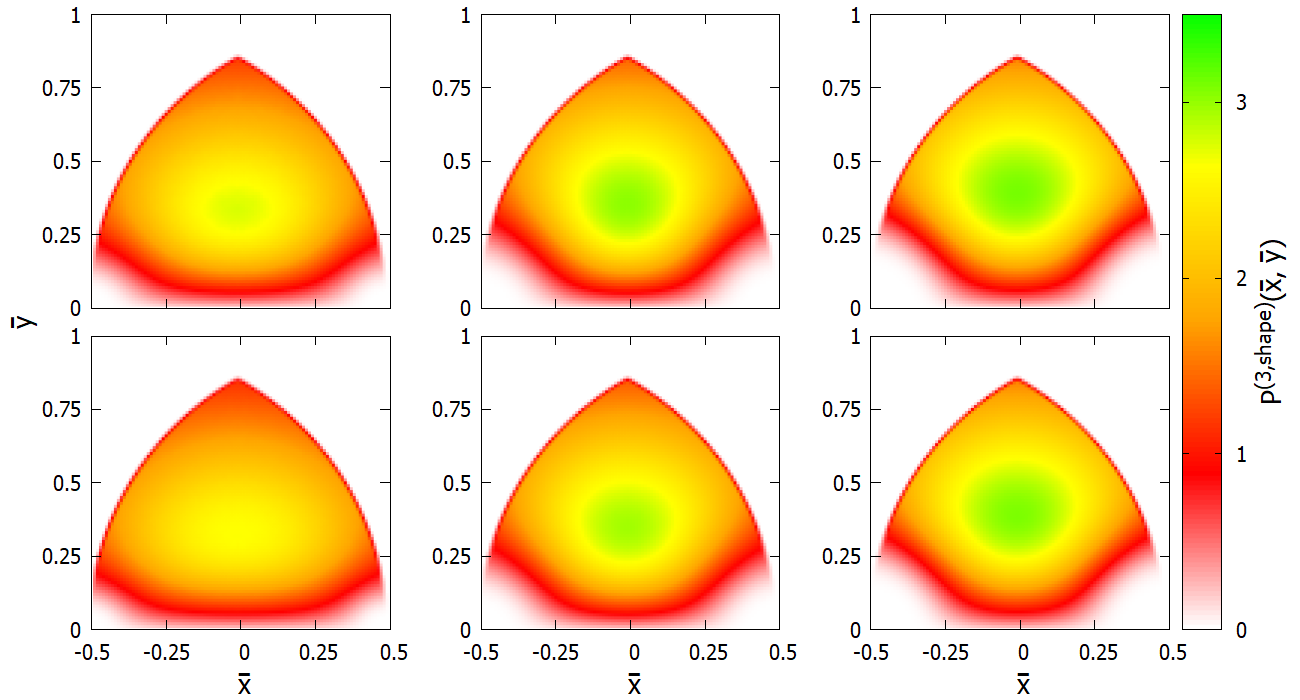}
\caption{$P_N^{(3,\text{shape})}(\bar{x},\bar{y})$ for Model~IB   at the physical point (top
row) and
at unitarity (bottom row).
The first, second, and third columns show
results for $N=3$, $N=4$, and $N=10$, respectively.
The color bar on the right applies to all six panels.
Since the triangles are oriented and normalized such that one particle sits
at $(\bar{x},\bar{y})=(-1/2,0)$ and the other at
$(\bar{x},\bar{y})=(+1/2,0)$ (with the interparticle distance vector corresponding to the largest distance being oriented along the
$\pm \bar{x}$-axis), the 
regions in the top left and top right of the panels are excluded by construction.
}
\label{fig_shape}
\end{figure}    

\end{widetext}

\begin{widetext}

\begin{figure}
\includegraphics[width=0.9\textwidth]{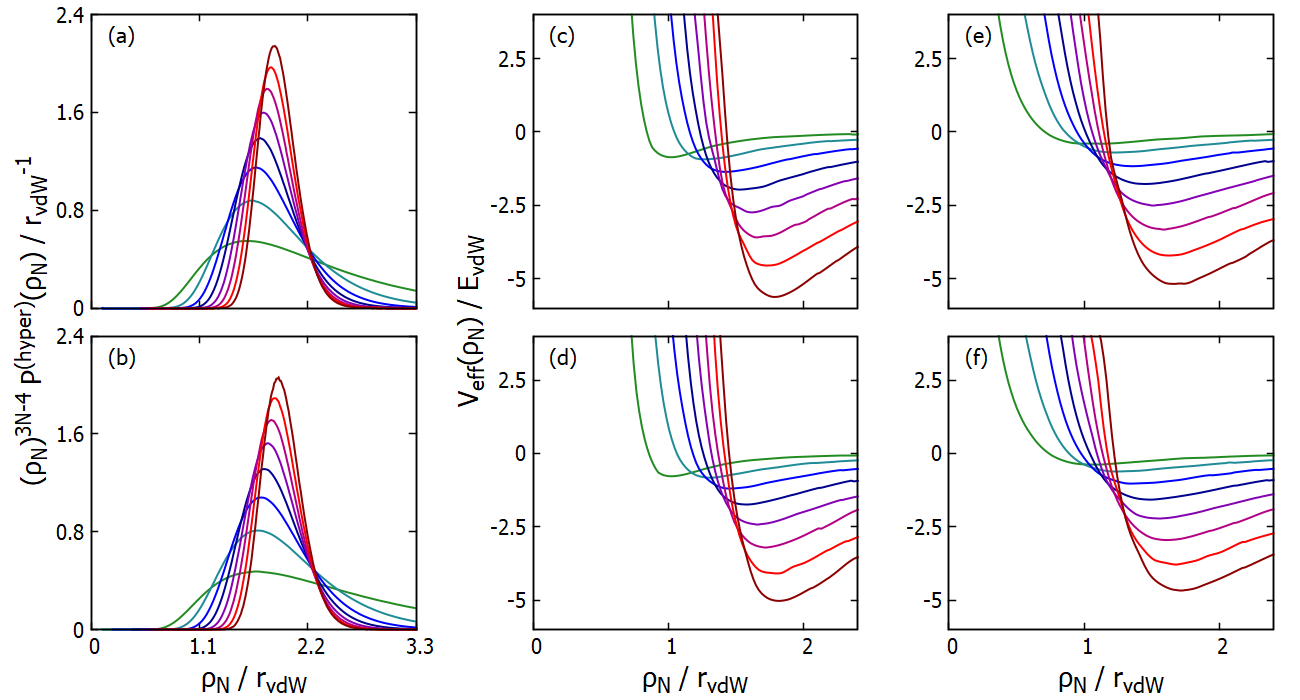}
\caption{Hyperradial properties 
for $N=3-10$  at the physical point (top
row) and
at unitarity (bottom row).
The first column shows $P_N^{(\text{hyper})}(\rho_N)$ for Model~IA.
The second column shows the (approximate) effective potential curves
$V_{\text{eff}}(\rho_N)$ for Model~IA,
calculated using Eq.~(\ref{equation_diff_hyper}).
For comparison, the third column shows $V_{\text{eff}}(\rho_N)$
for Model~II (the Model~II plots are made using the van der Waals length for Model~IA as a scale).
The color scheme is the same as in Fig.~\ref{fig_jacobi}.
}
\label{fig_hyper}
\end{figure}    

\end{widetext}

Figures~\ref{fig_hyper}(a) and \ref{fig_hyper}(b)
show the scaled hyperradial density $(\rho_N)^{3N-4} P_N^{(\text{hyper})}(\rho_N)$ 
for  $N=3-10$
clusters interacting through the 
CPKMJS potential
at the physical point and at unitarity, respectively.
The differences between the scaled hyperradial densities at the physical point and at unitarity for fixed $N$ are small. Careful inspection shows that the 
scaled hyperradial densities at unitarity extend to larger $\rho_N$ and rise at slightly larger $\rho_N$ than those at the physical point. Correspondingly, the maximum of 
$(\rho_N)^{3N-4} P_N^{(\text{hyper})}(\rho_N)$  is located at slightly larger $\rho_N$ for the clusters at unitarity than 
for the clusters at the physical point.
The fact that the scaled hyperradial densities at unitarity extend to larger $\rho_N$ than those at the physical point is 
a consequence of the smaller binding energy at unitarity than at the physical point. 
As $N$ increases, the scaled hyperradial densities become more localized, with their maximum shifting to larger $\rho_N$. To
interpret this behavior,
one needs to keep in mind that the definition of the hyperradius is intimately linked to the 
definition of the hyperradial mass $M$.
Since the
quantity $M \rho_N^2$ is an invariant but not $\rho_N$ and $M$ separately, 
$\rho_N$ can be multiplied by an overall 
factor~\cite{LIN19951,doi:10.1063/1.481404,D_Incao_2018}.
If all interparticle distances were equal to $\bar{r}$, then $\rho_N$ [as defined 
in Eq.~(\ref{eq_hyperradius})] would approach $\bar{r}/ \sqrt{2}$ in the $N \rightarrow \infty$ limit.
Since helium clusters behave roughly as incompressible liquids, the maximum of the hyperradial density is expected to 
occur at increasingly larger $\rho_N$ as $N$ increases.
Figures~\ref{fig_hyper}(a) and \ref{fig_hyper}(b) confirm this notion.

We now use the hyperradial densities to calculate approximate hyperradial
potential curves.
We note that Ref.~\cite{doi:10.1063/1.481404} obtained the effective hyperradial potential curves 
of helium clusters with $N=3-10$ at the physical point following an alternative and more rigorous approach;
in addition, Ref.~\cite{doi:10.1063/1.481404}  presented careful benchmark calculations
of the different approaches for $N=3$.
The approach pursued here yields potential curves that agree semi-quantitatively with 
%those 
the more accurate potential curves presented
in Ref.~\cite{doi:10.1063/1.481404}.
If the hyperradial and hyperangular degrees of freedom separate,
the effective one-dimensional 
Schroedinger equation for the lowest effective hyperradial potential curve
$V_{\text{eff}}(\rho_N)$ can be written in terms of
$F_N(\rho_N)$~\cite{Castin04,PhysRevA.74.053604,PhysRevLett.89.250401,PhysRevA.90.052514,PhysRevA.86.053633},
\begin{eqnarray}
\label{equation_diff_hyper}
\left[ 
-\frac{\hbar^2}{2M} \frac{\partial^2}{\partial \rho_N^2} + V_{\text{eff}}(\rho_N)
\right] F_N(\rho_N) = E_N F_N(\rho_N),
\end{eqnarray}
where
$F_N(\rho_N)=  [(\rho_N)^{3N-4} P_N^{(\text{hyper})}(\rho_N)]^{1/2}$.
For the $N$-particle clusters ($N \ge 3$) at unitarity, the separability is broken due to the finite-range nature
of the two-body interactions. 
At the physical point, the finiteness of the scattering length provides an additional 
separability-breaking mechanism. 
Even though Eq.~(\ref{equation_diff_hyper}) is not strictly valid for the 
potential models considered in this work, 
we ``invert'' it to obtain approximate effective hyperradial potentials
$V_{\text{eff}}(\rho_N)$.
The same strategy was pursued in Ref.~\cite{PhysRevA.90.052514} for $N=3$ and $4$.
Figures~\ref{fig_hyper}(c) and \ref{fig_hyper}(d)
show the results for Model~IB at the physical point and at unitarity,
respectively. The differences between the potential curves
at the physical point and at unitarity are very small.
Reference~\cite{PhysRevA.90.052514} conjectured, based on results for $N=3$ and $N=4$,
that the location of the repulsive inner wall of the hyperradial potential curves varies as 
$(N-1) r_{\text{vdW}}/ \sqrt{2N}$; this scaling accounts for an effective non-trivial reduction of the
configuration space due to an energy cost associated with adiabatic
deformation~\cite{PhysRevLett.112.105301,PhysRevA.90.022106}. This scaling 
was contrasted with an alternative scaling of $\sqrt{N-1} r_{\text{vdW}}/ \sqrt{2N}$,
which arises assuming that the minimum average interparticle spacing is given by $r_{\text{vdW}}$.
For $N=10$,  the inner wall would be located, according to these two scalings, at $0.671 r_{\text{vdW}}$
and $2.01 r_{\text{vdW}}$.
Figures~\ref{fig_hyper}(c) and \ref{fig_hyper}(d) show that the scaling is somewhere in between.

The corresponding effective hyperradial potential curves for Model~II
are shown in Figs.~\ref{fig_hyper}(e) and \ref{fig_hyper}(f). The effective potential curves for Model~II
are significantly softer (less steep) at small $\rho_N$ ($\rho_N/r_{\text{vdW}}$ between 
about $0.6$ and $1.2$) than those for Model~IB; this is consistent with what was discussed above for 
the pair distribution functions.

\begin{widetext}

\begin{figure}
\includegraphics[width=0.9\textwidth]{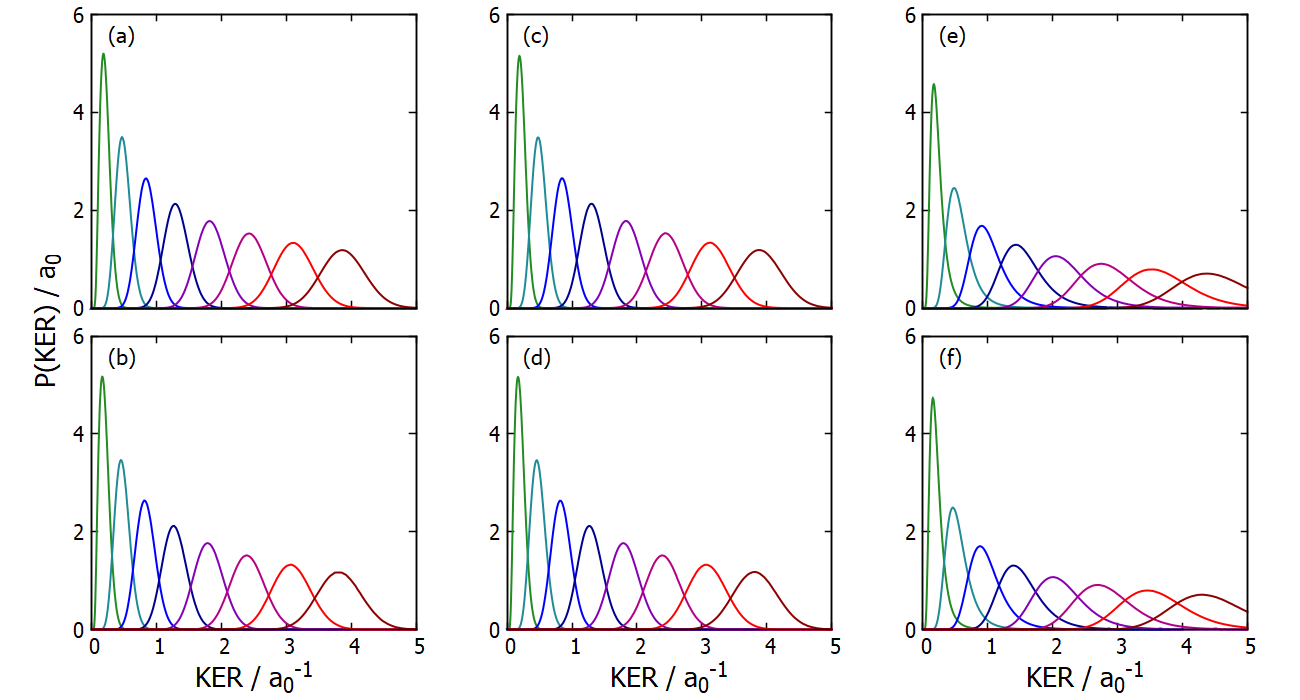}
\caption{KER
for $N=3-10$  at the physical point (top
row) and
at unitarity (bottom row).
The first, second, and third columns show the KER for
Model~IA, Model~IB, and Model~II, respectively.
The color scheme is the same as in Fig.~\ref{fig_jacobi}.
}
\label{fig_ker}
\end{figure}

\end{widetext}

Last, Fig.~\ref{fig_ker} presents the KER distribution functions at the physical point
(top row) and at unitarity (bottom row).
While small helium clusters  have been isolated in
molecular beam experiments~\cite{Voigtsberger2014,doi:10.1126/science.aaa5601}, 
Coulomb explosion experiments for $N \gtrsim 4$ are expected to be 
complicated by the fact the ions leaving the helium clusters might be undergoing additional 
collisions~\cite{Ulrich2011,PhysRevA.98.050701}. 
Despite of this challenge, we find it useful to analyze the dependence of the KER distribution functions
on the various interaction models. 
Since the number of interparticle distances increases as $N^2$ with increasing $N$,
the KER distribution functions move to larger KER with increasing $N$.
The KER distribution functions for Model~IA (first column) and Model~IB (second column)
are nearly indistinguishable on the scale shown. Careful inspection reveals small differences 
between the KER distribution functions  of clusters interacting through realistic interaction potentials
at the physical point and at unitarity.

The KER distribution functions for clusters interacting through Model~II extend to significantly larger
KER; this behavior is linked  to the enhanced probability for clusters interacting through Model~II, relative
to those interacting through realistic interaction potentials (Model~I), to find two particles at small interparticle distances. 
The broader KER distribution functions for Model~II also lead to peak values of the
KER distribution functions 
compared to those for Model~I.
We note that the KER distribution functions for Model~II do not only differ in the tail region from those for Model~I 
(high-energy region or short-distance region) but also in the ``rising portion'' of the KER distribution 
function (large distance region); these deviations are more pronounced for larger $N$ than for smaller $N$.
The deviations arise because the KER distribution functions
in the rising portion are not dominated by configurations in which all interparticle distances are large but by
configurations where $N-1$ interparticle distances are large and the remaining $N(N-1)/2-(N-1)$ 
interparticle distances are not particularly large.

\section{Conclusions}
\label{sec_conclusion}

This paper presented a comprehensive study of the structural properties of small 
bosonic helium clusters consisting of up to $N=10$ atoms and interacting through realistic
interaction potentials. In addition to helium clusters at the physical point,
characterized by a two-body $s$-wave scattering length that is positive and finite (and
notably larger than the van der Waals length), clusters interacting with an infinite $s$-wave scattering were investigated.
To reach unitarity, the realistic helium-helium interaction potential was multiplied
by an overall factor that is close to but smaller than one. 

For comparison, the properties of the systems at the physical point
and at unitarity were also calculated for an effective low-energy interaction model 
that was introduced in the literature~\cite{PhysRevA.102.063320}.
The model's 
strictly attractive two-body potential reproduces the two-body $s$-wave scattering length
and two-body binding energy obtained for the HFD-HE2 potential. A  strictly repulsive three-body potential is added to 
reproduce the three- and four-body energies obtained
for the HFD-HE2 potential.
Importantly, there is a difference between the effective low-energy model construction for clusters
at the physical point and at unitarity.
At the physical point, the requirements for matching the  two-body $s$-wave scattering length
and two-body binding energy are two distinct requirements.
At unitarity, in contrast, the two requirements are equivalent, i.e., fulfilling one of these requirements implies
that the other requirement is fulfilled automatically.

A detailed analysis of the structural properties at small and large length scales was presented,
with focus on comparing the results for different realistic interaction potentials and those for
the HFD-HE2 potential and the effective low-energy model.
Several small distance behaviors were found to be described accurately 
by the two-body correlation function for a purely attractive $-C_6/r^6$ potential.
The small length scale behavior of the pair distribution functions
for the realistic interaction models at the physical point was summarized by the 
two-body contact and the $(2+1)$ contact for each cluster size. 
The two-body contacts for different realistic interaction potentials
were found to be related to each other through, roughly, $N$-independent scaling factors. 
Following the spirit of Ref.~\cite{PhysRevA.101.010501},
it would be interesting to extend the current study to larger clusters and to extract, using the liquid drop model, the 
bulk pair-atom contact both at the physical point and at unitarity. It would also be interesting to investigate mixed
clusters that contain bosonic $^4$He and fermionic $^3$He atoms.

{\em{Acknowledgement:}}
Support by the National Science Foundation through
grant numbers PHY-1806259 and
PHY-2110158 is
gratefully acknowledged.
Work during the early stage was additionally supported by grant number NSF-1659501.
This work used
the OU
Supercomputing Center for Education and Research
(OSCER) at the University of Oklahoma (OU).

%\bibliography{helium_cluster}

\begin{thebibliography}{98}
\bibitem[{typ({})}]{typo_gaussian}
\bibinfo{journal}{{We note that the definition of the model in
  Ref.~\protect\cite{PhysRevA.102.063320} contains a typo:
  Reference~\protect\cite{PhysRevA.102.063320} reports $w_0$ as being positive;
  it is, however, negative. Our $R_0$ is related to $\rho_0$ through $\rho_0^2
  = 12 R_0^2$. }}  (\bibinfo{year}{{}}).
\bibitem[{eff({})}]{effective_range}
\bibinfo{journal}{{For the interaction potentials of Model~I-type, the
  effective range $r_{\text{eff}}$ is, in the large $|a_s|$ regime, roughly
  three times as large as the van der Waals length $r_{\text{vdW}}$.}}
  (\bibinfo{year}{{}}).
\end{thebibliography}

\end{document}